\documentclass[runningheads]{llncs}
\usepackage[T1]{fontenc}
\usepackage{graphicx}
\usepackage{hyperref}
\usepackage{color}
\usepackage{amsfonts}

\usepackage{cite}
\usepackage{lineno}

\usepackage{xcolor}
\definecolor{clr-background}{RGB}{255,255,255}
\definecolor{clr-background-pvl}{RGB}{245,245,245}
\definecolor{clr-text}{RGB}{0,0,0}
\definecolor{clr-string}{RGB}{0, 128, 0}
\definecolor{clr-variable}{RGB}{0,0,0}
\definecolor{clr-comment}{RGB}{64, 128, 128}
\definecolor{clr-preprocessor}{RGB}{128,128,128}
\definecolor{clr-keyword}{RGB}{0, 128, 0}
\definecolor{clr-type}{RGB}{0, 0, 255}
\definecolor{clr-class}{RGB}{0, 0, 255}
\definecolor{clr-operator}{RGB}{10,10,10}
\definecolor{clr-number}{RGB}{0,0,0} 
\definecolor{clr-constant}{RGB}{136, 0, 0} 
\usepackage{multirow}

\definecolor{ao}{rgb}{0.0, 0.3, 0.0}
\definecolor{burgundy}{rgb}{0.5, 0.0, 0.13}
\definecolor{ceruleanblue}{rgb}{0.10, 0.30, 0.90}

\usepackage{ stmaryrd }
\usepackage{listings}
\usepackage{paralist}
\usepackage{float}
\newfloat{lstfloat}{htbp}{lop}
\floatname{lstfloat}{Listing}
\newcommand*{\ditto}{\textquotedbl}

\lstdefinestyle{halide}{
  float=t,
  breaklines,
  language=C++,
  backgroundcolor=\color{clr-background},
  basicstyle=\ttfamily\color{clr-text}\lst@ifdisplaystyle\scriptsize\fi,
  stringstyle=\color{clr-string},
  identifierstyle=\color{clr-variable},
  commentstyle=\it\color{clr-comment},
  directivestyle=\color{clr-preprocessor},
  keywordstyle=\bf\color{clr-keyword},
  numbers=left,
  tabsize=2,
  columns=fullflexible,
  xleftmargin=1mm,
  alsoletter=\,*,
  frame = single,
  framesep=\fboxsep,
  framerule=\fboxrule,
  mathescape=true,
  showstringspaces=false,
  showspaces=false,
  xleftmargin=0.05\textwidth,
  xrightmargin=0.01\textwidth,
  classoffset=1,
  morekeywords={Func,Var,Exp,RDom,Buffer,result, forall, exists, gtid, ltid, pipeline, fun, allocate, label, rdom, select, produce,store,split,compute_at,store_at,unroll,parallel,serial fuse,fuse,reorder,clone_in,unrolled,consume,vectorize},
  keywordstyle=\color{clr-class},
  classoffset=2,
    literate=%
    {0}{{{\color{clr-number}0\lst@whitespacefalse}}}1
    {1}{{{\color{clr-number}1\lst@whitespacefalse}}}1
    {2}{{{\color{clr-number}2\lst@whitespacefalse}}}1
    {3}{{{\color{clr-number}3\lst@whitespacefalse}}}1
    {4}{{{\color{clr-number}4\lst@whitespacefalse}}}1
    {5}{{{\color{clr-number}5\lst@whitespacefalse}}}1
    {6}{{{\color{clr-number}6\lst@whitespacefalse}}}1
    {7}{{{\color{clr-number}7\lst@whitespacefalse}}}1
    {8}{{{\color{clr-number}8\lst@whitespacefalse}}}1
    {9}{{{\color{clr-number}9\lst@whitespacefalse}}}1
    {\\forall}{{{\color{clr-operator}$\forall$\lst@whitespacefalse}}}1
    {<}{{{\color{clr-operator}$<$\lst@whitespacefalse}}}1
    {<=}{{{\color{clr-operator}$\leq$\lst@whitespacefalse}}}1
    {==}{{{\color{clr-operator}$ == $\lst@whitespacefalse}}}1
    {\&\&}{{{\color{clr-operator}$\wedge$\lst@whitespacefalse}}}1
    {||}{{{\color{clr-operator}$\vee$\lst@whitespacefalse}}}1
    {>}{{{\color{clr-operator}$>$\lst@whitespacefalse}}}1
    {>=}{{{\color{clr-operator}$\geq$\lst@whitespacefalse}}}1
    {==>}{{{\color{clr-operator}$\Rightarrow$\lst@whitespacefalse}}}1
    {=>>}{{{\color{clr-operator}$\mapsto$\lst@whitespacefalse}}}1
    {!=}{{{\color{clr-operator}$\neq$\lst@whitespacefalse}}}1,
  otherkeywords={.,;,-,+,*,~},
  morekeywords={,.,;,-,+,*,/,~},
  keywordstyle={\color{clr-operator}},
  classoffset=0,
  morekeywords={requires, ensures, loop_invariant, loop, invariant, invariant_perm, context,context_perm, in,pipeline_requires,pipeline_ensures},
  emph={int,char,double,float,unsigned,void,bool},
  emphstyle=\color{clr-type},
}

\lstdefinestyle{pvl}{
float=t,
language = C++,
numbers=left,
commentstyle=\color{ceruleanblue},
moredelim = *[s][]{/*@}{@*/},
moredelim = *[l][]{//@},
backgroundcolor=\color{clr-background-pvl},
columns=fullflexible,
xleftmargin=1mm,
alsoletter=\,*,
morekeywords={pure, invariant, ensures, requires, context, decreases, context_everywhere, \old, \length, par, \result,
  \forall, \forall*, **, Perm,send,receive,__kernel,loop_invariant,for, Barrier, barrier, resource, CLK_GLOBAL_MEM_FENCE, get_global_id, seq, frac, inline, tail, head, none, int32_t, int64_t, pragma}, 
breaklines=true,
tabsize=2,
basicstyle=\ttfamily\lst@ifdisplaystyle\scriptsize\fi,
keywordstyle=\color{burgundy}\tt,
frame = single,
framesep=\fboxsep,
framerule=\fboxrule,
mathescape=true,
showstringspaces=false,
showspaces=false,
xleftmargin=0.05\textwidth,
xrightmargin=0.01\textwidth,
literate=%
    {0}{{{\color{clr-number}0\lst@whitespacefalse}}}1
    {1}{{{\color{clr-number}1\lst@whitespacefalse}}}1
    {2}{{{\color{clr-number}2\lst@whitespacefalse}}}1
    {3}{{{\color{clr-number}3\lst@whitespacefalse}}}1
    {4}{{{\color{clr-number}4\lst@whitespacefalse}}}1
    {5}{{{\color{clr-number}5\lst@whitespacefalse}}}1
    {6}{{{\color{clr-number}6\lst@whitespacefalse}}}1
    {7}{{{\color{clr-number}7\lst@whitespacefalse}}}1
    {8}{{{\color{clr-number}8\lst@whitespacefalse}}}1
    {9}{{{\color{clr-number}9\lst@whitespacefalse}}}1
    {\\forall}{{{\color{clr-operator}$\forall$\lst@whitespacefalse}}}1
    {<}{{{\color{clr-operator}$<$\lst@whitespacefalse}}}1
    {<=}{{{\color{clr-operator}$\leq$\lst@whitespacefalse}}}1
    {==}{{{\color{clr-operator}$ \equiv $\lst@whitespacefalse}}}1
    {\&\&}{{{\color{clr-operator}$\wedge$\lst@whitespacefalse}}}1
    {||}{{{\color{clr-operator}$\vee$\lst@whitespacefalse}}}1
    {>}{{{\color{clr-operator}$>$\lst@whitespacefalse}}}1
    {>=}{{{\color{clr-operator}$\geq$\lst@whitespacefalse}}}1
    {==>}{{{\color{clr-operator}$\Rightarrow$\lst@whitespacefalse}}}1
    {=>>}{{{\color{clr-operator}$\mapsto$\lst@whitespacefalse}}}1
    {->}{{{\color{clr-operator}$\shortrightarrow$\lst@whitespacefalse}}}1
    {!=}{{{\color{clr-operator}$\neq$\lst@whitespacefalse}}}1,
}

\def\halinline{\lstinline[style=halide]}
\def\pvlinline{\lstinline[style=pvl]}

\newcommand{\hoare}[1]{\big\{ #1 \big\}}

\usepackage{proof}
\usepackage{syntax}

\usepackage{tikz}
\usepackage[utf8]{inputenc}

\usepackage{pmboxdraw}

\newcommand{\vbar}{|}
\lstMakeShortInline[columns=fixed, breaklines]|

\usepackage{xspace}

\newcommand{\halide}{\textsc{Halide}\xspace}
\newcommand{\haliver}{\textsc{HaliVer}\xspace}
\newcommand{\vercors}{\textsc{VerCors}\xspace}
\newcommand{\coq}{Coq\xspace}
\newcommand{\pvl}{\textsc{Pvl}\xspace}
\newcommand{\C}{\textsc{C}\xspace}
\newcommand{\cogent}{\textsc{Cogent}\xspace}
\newcommand{\isabellehol}{\textsc{Isabelle/HOL}\xspace}
\newcommand{\compcert}{\textsc{CompCert}\xspace}
\newcommand{\alpinist}{\textsc{Alpinist}\xspace}

\newcommand{\viper}{\textsc{Viper}\xspace}

\def\CC{{C\nolinebreak[4]\hspace{-.05em}\raisebox{.4ex}{\tiny\bf ++}}\xspace}

\newcommand{\beginremove}{\color{orange}}

\begin{document}
%
\title{\haliver: Deductive Verification and Scheduling Languages Join Forces}

%
\makeatletter
\newif\if@remove
\@removetrue

\newif\if@removetwo
\@removetwofalse

\newcommand{\remove}[1]{\if@remove\else{\color{orange}#1}\fi}
\newcommand{\add}[1]{\if@remove#1\else\fi}
\newcommand{\removetwo}[1]{\if@removetwo\else{#1}\fi}
\newcommand{\addtwo}[1]{\if@removetwo#1\else\fi}

\newif\if@blind
\@blindfalse 
\if@blind 
\author{Anonymous\inst{1}}\institute{Anonymous Institute}
\else
\author{Lars B. van den {Haak}\inst{1}\orcidID{0000-0002-0330-5016} \and
Anton {Wijs}\inst{1}\orcidID{0000-0002-2071-9624} \and
Marieke {Huisman}\inst{2}\orcidID{0000-0003-4467-072X}
\and
Mark van den {Brand}\inst{1}\orcidID{0000-0003-3529-6182}}
\authorrunning{L. B. van den Haak, A.J. Wijs, M. Huisman and M.G.J. van den Brand}
\institute{
Eindhoven University of Technology, The Netherlands \\
\email{\{l.b.v.d.haak, a.j.wijs, m.g.j.v.d.brand\}@tue.nl} \and
University of Twente, The Netherlands \\
\email{m.huisman@utwente.nl}
}
\fi
\maketitle              
\begin{abstract}
	The \haliver tool integrates deductive verification into the popular scheduling language \halide, used for image processing pipelines and array computations. \haliver uses \vercors, a separation logic-based verifier, to verify the correctness of (1) the \halide algorithms and (2) the optimised parallel code produced by \halide when an optimisation schedule is applied to an algorithm. This allows proving complex, optimised code correct while reducing the effort to provide the required verification annotations. For both approaches, the same specification is used.	
	We evaluated the tool on several optimised programs generated from characteristic \halide algorithms, using all but one of the essential scheduling directives available in \halide. Without annotation effort, \haliver proves memory safety in almost all programs. With annotations \haliver, additionally, proves functional correctness properties.	    
	We show that the approach is viable and reduces the manual annotation effort by an order of magnitude.
	\keywords{Program correctness  \and Deductive verification \and Scheduling language.}
\end{abstract}


\section{Introduction}\label{sec:introduction}

To meet the continuously growing demands on software performance, parallelism is increasingly often needed~\cite{Leisersoneaam9744}. However, introducing parallelism tends to increase the risk of introducing errors, as the interactions between parallel computations can be hard to predict. Moreover, a plethora of optimisation techniques exists~\cite{HijHe23}, so identifying when an optimisation can be applied safely, without breaking correctness, can be very challenging. Also, applying optimisations tends to make a program more complex, making it harder to reason about.

To address this, on the one hand, various domain-specific languages (DSLs) have been proposed that separate the \textit{algorithm} (\emph{what} it does) from the parallelisation \textit{schedule} (\emph{how} it does it). These are called \emph{scheduling languages}~\cite{ragan-kelleyHalideDecouplingAlgorithms2017,ragan-kelleyHalideLanguageCompiler2013,chenTVMAutomatedEndtoEnd2018,tiramisu,chill,graphit,hagedornFireironDataMovementAwareScheduling2020}. Given an algorithm and a schedule, a compiler generates an optimised parallel program. This approach crucially depends on the schedule not introducing any errors in the functionality, which is not always obvious.

On the other hand, \emph{deductive program verification}~\cite{deductivever} has been successfully applied to verify the functionality of parallel programs~\cite{blomVerCorsToolSet2017}. This requires that the intended functionality is formalised as a contract, for instance using \emph{permission-based separation logic}~\cite{AmighiHHH15,Bornat05}. A major hurdle, preventing this technique from being adopted at a large scale, is that if a program becomes more complicated, the required annotations rapidly
grow in size and complexity~\cite{safari2020formal,safariFormalVerificationParallel2020}.


\begin{figure}[t]
\centering
\scalebox{0.6}{

\usetikzlibrary{
  shapes.arrows,
  arrows,
  arrows.meta,                        
  backgrounds,                        
  calc}                               
\tikzset{
  basic box/.style={
    shape=rectangle, rounded corners, align=center, draw=#1, fill=#1!25
    , font=\normalsize, text depth=+.3ex},
  text/.style={ font=\small},
}

\begin{tikzpicture}[]
  \draw node[basic box=blue, minimum width = 2cm, minimum height = 1cm, outer sep=0.5cm] (hal) at (0,0) {\halide Algorithm} ;
  \draw node[basic box=green, minimum width = 2cm, minimum height = 1cm, outer sep=0.5cm] (hal-ann) at ([yshift=-0.1cm]hal.south) {Annotations} ;


  \draw node[basic box=yellow, minimum width = 2cm, minimum height = 1cm, outer sep=0.5cm] (par) at ([xshift=3.2cm]hal.east) {Optimised \C Program} ;
  \draw node[basic box=green, minimum width = 2cm, minimum height = 1cm, outer sep=0.2cm] (par-ann) at ([yshift=-0.1cm]par.south)  {Annotations Matching\\ Optimised \C Program} ;

  \node[single arrow, draw, minimum width = 0.2cm, minimum height=1.5cm] (arrow-right) at ([yshift=-0.5cm, xshift=0.4cm] hal.east) {};
  \node[text=] at ([yshift=0.4cm]arrow-right.north) (backtext) {Back-end};
  \draw node[basic box=blue, minimum width = 1cm, minimum height = 1cm, outer sep=0.2cm] (sched) at ([yshift=-1.0cm]arrow-right.south)  {Schedule} ;
  \draw ([yshift=-0.1cm]arrow-right.south) edge node[]{\textbf{+}} (sched);

  \node[single arrow, draw, minimum width = 0.2cm, minimum height=1.5cm,rotate=180] (arrow-left) at ([yshift=-0.5cm, xshift=-0.4cm] hal.west) {};
  \node[text=] at ([yshift=0.4cm]arrow-left.south) {Front-end};

  \draw node[basic box=yellow, minimum width = 2cm, minimum height = 1cm, outer sep=0.5cm] (pvl) at ([xshift=-3.2cm]hal.west)  {\vercors Encoding} ;
  \draw node[basic box=green, minimum width = 2cm, minimum height = 1cm, outer sep=0.2cm] (pvl-ann) at ([yshift=-0.1cm]pvl.south)  {\vercors Annotations} ;

  \node at ([yshift=-1.0cm, xshift=-1.5em]pvl-ann.south) (accept-1) {\Large$\textcolor{teal}\checkmark$};
  \node at ([xshift=3em]accept-1)  (reject-1) {\Large\textcolor{red}X};

  \draw [->](pvl-ann.south) -- ([xshift=1.5em]accept-1.north);
  \node[text=] at ([xshift=+3em,yshift=-0.5cm]pvl-ann.south) {\vercors};

  \node at ([yshift=-1.0cm, xshift=-1.5em]par-ann.south) (accept-2) {\Large$\textcolor{teal}\checkmark$};
    \node at ([xshift=3em]accept-2)  (reject-2) {\Large\textcolor{red}X};

    \draw [->](par-ann.south) -- ([xshift=1.5em]accept-2.north);
    \node[text=] at ([xshift=+3em,yshift=-0.5cm]par-ann.south) {\vercors};

\end{tikzpicture}}
\caption{\label{fig:process} High level overview of our approach.}
\end{figure}
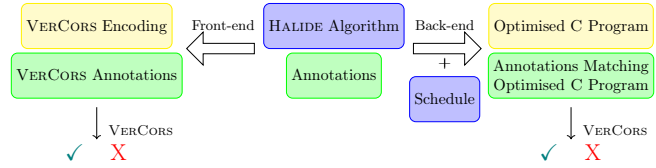

In this paper, we combine the best of both worlds. 
We propose the \haliver tool, which focusses
on \halide~\cite{ragan-kelleyHalideDecouplingAlgorithms2017,ragan-kelleyHalideLanguageCompiler2013}, a scheduling language for portable image computations and array processing. It has been widely adopted in industry, for instance to produce parts of Adobe Photoshop and to implement the YouTube video-ingestion pipeline. For verification, we use the \vercors program verifier~\cite{blomVerCorsToolSet2017}. 
In this paper we define two verification approaches (1) \textit{front-end} and (2) \textit{back-end}, as seen in Figure~\ref{fig:process}. Our approaches verify that the program adheres to the \textit{same} functional specification. This specification is detailed by annotating the algorithmic part of a \halide program, thereby keeping the annotations focussed on the functionality, and therefore relatively straightforward.
With the \textit{front-end} verification approach we verify the correctness of the algorithmic part of a \halide program. \haliver transforms the algorithm and the annotations to an annotated \vercors program.
With \textit{back-end} verification approach we verify the \C code that the \halide compiler generates, given a \halide algorithm and a schedule. \haliver transforms the given annotations to match the generated code.
Furthermore, where possible, \haliver generates annotations, such as permission specifications, to relieve the user from having to manually write these. This contributes to making the annotation process straightforward.  

In this way, \haliver allows the user to succinctly specify the intended functionality of optimised, parallel code, and it checks that the resulting program indeed has the desired functionality. A major advantage of our approach is that it is flexible to use in a setting where multiple compiler passes are made. Also, it can be easily extended if a new compiler pass or schedule optimisation is added. An alternative would be to prove correctness of the compiler, but this would require a large amount of initial work and additionally for each change to the compiler.


Concretely, this paper provides the following contributions:
\begin{compactitem}
\item An annotation language to describe the functionality of \halide algorithms, which is integrated into the \halide algorithm language;
\item Tool support for the front-end verification approach of \halide algorithms; 
\item Tool support for the back-end verification approach, which can verify programs generated by the \halide compiler from an algorithm and a schedule;
\item Evaluation of the \haliver tool on \halide examples using all but one of the essential scheduling directives available in various combinations.
	We evaluated the tool on 23 different optimised programs, generated from eight characteristic \halide algorithms, to prove memory safety with no annotation effort. For 21 cases, \haliver proves safety, for the remaining two cases we discuss the limitations.
	For 20 programs, based on five algorithms, we also add annotations for functional correctness properties. For 19 of these programs \haliver proves correctness, for the remaining one we run into a similar limitation.

\end{compactitem}

The remainder of this paper is organised as follows.
Section~\ref{sec:background} gives brief background information on \halide and \vercors. Section~\ref{sec:approach} introduces \halide annotations, and describes how \haliver supports the verification of an algorithm
and an optimised program. The approach is illustrated on characteristic examples. Section~\ref{sec:experiments} evaluates the \haliver tool, and Sections~\ref{sec:related-work} and~\ref{sec:Discussion} address related work, conclusions and future work.


\section{Background}\label{sec:background}

\paragraph{\textbf{\halide.}}
\halide is a DSL embedded in \CC, targeting image processing pipelines and array computations~\cite{ragan-kelleyHalideDecouplingAlgorithms2017,ragan-kelleyHalideLanguageCompiler2013}.\footnote{A \halide tutorial can be found here: \url{https://halide-lang.org/tutorials/}.} \halide separates the \textit{algorithm}, defining what you want to calculate, from the \textit{schedule}, defining how the calculation should be performed. Typically, when optimising code for a specific architecture, the code becomes
much more complex and loses portability. By separating the schedule, the code expressing the functionality is not altered.

\if@remove
\begin{lstlisting}[style=halide, caption={\halide blur example with annotations added to verify the code.}, label=lst:blur-halide]
  requires inp.x.min == blur_y.x.min && inp.x.max == blur_y.x.max+2 && inp.y.min == blur_y.y.min && inp.y.max == blur_y.y.max+2;
  ensures \forall x, y . blur_y.x.min<=x<blur_y.x.max && blur_y.y.min<=y<blur_y.y.max ==> blur_y(x,y) == ((inp(x,y)+inp(x+1,y)+inp(x+2,y))/3 + (inp(x,y+1) +inp(x+1,y+1)+inp(x+2,y+1))/3 + (inp(x,y+2)+inp(x+1,y+2) + inp(x+2,y+2))/3)/3);
void blur(Buffer<2,int> inp, Func &blur_y){
  Func blur_x; Var x, y;
  blur_x(x,y) = (inp(x,y) + inp(x+1,y) + inp(x+2,y))/3
  blur_x.ensures(blur_x(x,y) == (inp(x,y) + inp(x+1,y) + inp(x+2,y))/3);
  blur_y(x,y) = (blur_x(x,y) + blur_x(x,y+1) + blur_x(x,y+2))/3;
  blur_y.ensures(blur_y(x,y) == ((inp(x,y)+inp(x+1,y)+inp(x+2,y))/3 + (inp(x,y+1)+inp(x+1,y+1) + inp(x+2,y+1))/3 + (inp(x,y+2)+inp(x+1,y+2)+inp(x+2,y+2))/3)/3;}
\end{lstlisting}
\else
\begin{lstlisting}[style=halide, caption={\remove{The functionality of a box filter (blur), expressed in \halide.}}, label=lst:blur]
void blur(Buffer<2,int> inp, Func &blur_y){
  Var x, y; Func blur_x;
  blur_x(x,y) = (inp(x,y) + inp(x+1,y) + inp(x+2,y))/3
  blur_y(x,y) = (blur_x(x,y) + blur_x(x,y+1) + blur_x(x,y+2))/3;}
\end{lstlisting}
\fi
\if@remove
Listing~\ref{lst:blur-halide} presents the \halide algorithm for a box filter, or blur function. The reader can ignore the \halinline|requires| and \halinline|ensures| annotations for now.
\else
Listing~\ref{lst:blur} presents the \halide algorithm for a box filter, or blur function.
\fi
Images are represented as pure (side-effect free) functions that point-wise map coordinates to \remove{image }values. A blur function defines how every pixel, referred to by its two-dimensional
coordinates, should be updated. In the example, the coordinates are
represented by
the variables \halinline|x| and \halinline|y|.
\halide uses a functional style, allowing algorithms
to be \remove{very }compact and loop-free. 
\halide functions are denoted by the keyword \halinline|Func|.
In the example, the input image is stored in a two-dimensional integer buffer \halinline|inp|,
and the output is given by defining the function \halinline|blur_y|, a reference to which is a parameter of \halinline|blur|. A \emph{pipeline} of function calls is defined: \remove{first, }the function \halinline|blur_x| is applied on the input image
\if@remove
(line 5).
\else
(line 3).
\fi
The output of that function is \remove{subsequently }used to compute the final image with the function \halinline|blur_y| 
\if@remove
(line 7).
\else
(line 4).
\fi
With \halinline|inp.x.min| and \halinline|inp.x.max| we refer to the minimum and maximum value of the dimension \halinline|inp.x|, respectively. 

A function may involve \emph{update definitions}, which (partially) update the value of a function. A \emph{reduction domain} is a way to apply an update a finite number of times and is typically used to express sums or histograms in \halide. A function is called a \emph{reduction} when such a domain is used, and an initialisation and an update definition are given. Listing~\ref{lst:reduction-sum} presents a reduction example.
For now, ignore the \halinline|ensures| and \halinline|invariant| lines. The reduction domain (\halinline|RDom|) \halinline|r| ranges from \halinline|0| to \halinline|9|, i.e. it consists of \halinline|10| values. The initial value of the \halinline|count|
function is defined at line 3, and line 5 is executed once for every value in \halinline|r|.
The statement \halinline|select(a,b,c)| returns \halinline|b| if \halinline|a| evaluates to \textbf{true}, \halinline|c| otherwise. For a given matrix of integers \halinline|inp|,
\halinline|cnt| counts the number of non-zeros at the first ten positions of each row in \halinline|inp|.

A \halide \textit{schedule} is given in Listing~\ref{lst:back-end-loopnest} and further explained in Section~\ref{sec:back-end}.

\begin{lstlisting}[style=halide, caption={A reduction to count the positive numbers of each row in matrix \halinline|inp|.}, label=lst:reduction-sum]
	void cnt(Buffer<2,int> inp, Func count) {
		Var x; RDom r(0,10);
		count(x) = 0;
		count.ensures(count(x) == 0);
		count(x) = select(inp(x, r) > 0, count(x)+1, count(x))
		count.invariant(0<=count(x)<=r);
		count.ensures(0<=count(x)<=10);}
\end{lstlisting}



\paragraph{\textbf{\vercors.}}
\vercors\footnote{An online tutorial can be found at \url{https://vercors.ewi.utwente.nl/wiki/}.}~\cite{blomVerCorsToolSet2017} is a deductive verifier to verify the functional correctness of, possibly concurrent, software.
Its specification language uses permission\hspace{0cm}-based separation logic~\cite{Bornat05}, a combination
of first-order logic and read/write permissions. The latter are used for concurrency-related verification, to express which data can be accessed by a thread at which moment.
Programs written in a number of languages, such as \textsc{Java} and \C, can be verified. \vercors also has
its own language, \pvl. \remove{\pvl is similar to Java and \CC, as it has classes and methods, but it also allows to express parallel blocks. }\vercors's verification engine relies on \viper~\cite{MullerSS16},
which applies symbolic execution to analyse programs with persistent mutable state.

Intended functional behaviour can be specified by means of pre- and postconditions, indicated by the keywords \pvlinline{requires} and \pvlinline{ensures},
respectively. The statement \pvlinline{context P} is an abbreviation for \pvlinline{requires P; ensures P}.
 Loop invariants and assertions can be added to the
code to help \vercors in proving the pre- and postconditions. We refer to the pre- and postconditions, loop invariants and assertions together as the \emph{annotations}
of a code fragment.
A permission \pvlinline|Perm(x, f)| gives permission to memory location \pvlinline|x|, where \pvlinline|f| is a fractional, with \pvlinline|1\1| indicating a write and anything between \pvlinline|0\1| and \pvlinline|1\1| a read. 
For a statement $s$, we have the Hoare triple $\hoare{P}s\hoare{Q}$. This indicates that if $P$ holds in the \textit{pre-state} then after $s$, $Q$ holds in the \textit{post-state}. %
A \textit{pure} function is without side-effects, thus can be used in annotations. It has the keyword \pvlinline{pure} in the header, and its body is a single expression. Annotations and pure function definitions in \C files are given in special comments, like \pvlinline|//@| or \pvlinline|/*@...@*/| for multi-line comments. (See Listing \ref{lst:blur-translation-consume-blur-x} for examples.)

\vercors can prove termination of recursive functions. Whenever the clause \pvlinline{decreases r} is added to a function contract, \vercors will try to prove that the function terminates, by showing that all recursive calls will strictly decrease the value of \pvlinline{r} while \pvlinline{r} has a lower bound.



\section{Verification of Scheduling Languages with \haliver }\label{sec:approach}
\haliver works directly on a \halide program and its intermediate representations, adding and transforming annotations where necessary. The tool is embedded in the \halide compiler. From a user's point of view, the general approach is as follows, using the front-end and back-end approach as in Figure~\ref{fig:process}.
\begin{enumerate}
\item \textbf{Write a \halide algorithm and add annotations}.
Annotations are the functional specification of the \halide algorithm.
Since a user can write an incorrect \halide algorithm, its correctness is ideally checked against a user-supplied specification.
\item \textbf{The front-end approach produces a \pvl encoding.} This encoding contains the algorithm and the specified annotations.
\item \textbf{\vercors verifies the encoding.} If verification succeeds, we know that the front-end algorithm conforms to the functional specification.
Otherwise, the verification fails; \vercors produces a counterexample and we return to step~1.
\item \textbf{Write a \halide schedule}.
\item \textbf{The back-end approach produces an annotated \C file.}  The tool automatically generates permission annotations. These allow us to prove data-race freedom and the absence of out-of-bound errors. The tool transforms the \remove{algorithm }annotations and generates additional annotations to match the scheduled back-end code. This is highly non-trivial, as each \remove{parallel block and }for-loop requires precise annotations to guide \vercors in the verification. However, it is ensured that the same property
is verified.
\item \textbf{\vercors verifies the back-end \C file.} If the verification fails, the lines of \C code that caused the failure are given, which can be traced back to the \halide algorithm. The cause of a verification failure may be that
\begin{itemize}
\item The \halide compiler produced incorrect code w.r.t. the specifications.
\item More \textit{auxiliary} annotations from step 1 are needed to guide \vercors.
\item A limitation has been encounter of the tools \haliver relies on, e.g., \vercors or the underlying SMT solver.
\end{itemize}
\end{enumerate}

In the remainder of this section we explain how to write annotations, and address front-end and back-end verification approaches. We also discuss the soundness and current limitations of the technique.

\subsection{\halide annotations}\label{sec:halide-ann}
\haliver makes it possible to add annotations when writing a \halide algorithm. Intuitively, these annotations are added as a Hoare triple. We consider three types of annotation: \textit{pipeline}, \textit{intermediate} and \textit{reduction invariant} annotations.


\if@remove

\else
\begin{lstlisting}[style=halide, caption={\halide blur example with annotations added to verify the code.}, label=lst:blur-halide]
  requires inp.x.min == blur_y.x.min && inp.x.max == blur_y.x.max+2 && inp.y.min == blur_y.y.min && inp.y.max == blur_y.y.max+2;
  ensures \forall x, y . blur_y.x.min<=x<blur_y.x.max && blur_y.y.min<=x<blur_y.y.max ==> blur_y(x,y) == ((inp(x,y)+inp(x+1,y)+inp(x+2,y))/3 + (inp(x,y+1) +inp(x+1,y+1)+inp(x+2,y+1))/3 + (inp(x,y+2)+inp(x+1,y+2) + inp(x+2,y+2))/3)/3);
void blur(Buffer<2,int> inp, Func &blur_y){
  Func blur_x; Var x, y;
  blur_x(x,y) = (inp(x,y) + inp(x+1,y) + inp(x+2,y))/3
  blur_x.ensures(blur_x(x,y) == (inp(x,y) + inp(x+1,y) + inp(x+1,y))/3);
  blur_y(x,y) = (blur_x(x,y) + blur_x(x,y+1) + blur_x(x,y+2))/3;
  blur_y.ensures(blur_y(x,y) == ((inp(x,y)+inp(x+1,y)+inp(x+2,y))/3 + (inp(x,y+1)+inp(x+1,y+1) + inp(x+2,y+1))/3 + (inp(x,y+2)+inp(x+1,y+2)+inp(x+2,y+2))/3)/3;}
\end{lstlisting}
\fi

In Listing~\ref{lst:blur-halide} annotations have been added. The lines 1--2 are \textit{pipeline annotations}: they specify the pre- and postconditions of the whole function and can only contain references to input buffers or output functions. Note that the results are stored directly in the \halinline|blur_y| function. Line 1 specifies how the input and output bounds should be related. Line 2 indicates what the output values are. One can add \textit{intermediate annotations} after any (update) function call to specify state predicates for particular locations in the pipeline. Examples are the \halinline|blur_x.ensures| and \halinline|blur_y.ensures| state predicates of Listing~\ref{lst:blur-halide} (lines 6 and 8).

\halide functions map coordinates to values pointwise. To achieve a one-to-one relationship between function and annotations, the intermediate annotations for a function should also specify how coordinates relate to values pointwise.
However, input buffers can be used freely with any point. For example, \halinline|blur_x.ensures(blur_x(x,y) >= inp(x+1,y))| is valid, but \halinline|blur_x.ensures(blur_x(x+1,y) >= 0)| is not, because the latter refers to \halinline|blur_x(x+1,y)| as opposed to \halinline|blur_x(x,y)|. \haliver requires this because each point of the function may be computed in parallel in the back-end, so it must be possible to reason about the points individually.

For ease of annotation, \haliver automatically generates a pipeline postcondition. This postcondition is derived from the
intermediate annotation of the last pipeline function in the algorithm. For Listing~\ref{lst:blur-halide}, \haliver can generate line 2, which is
 included here for completeness, based on line 8.

To prove that a \emph{reduction} is correct, \textit{reduction invariant} annotations must be provided for reduction domains. In Listing~\ref{lst:reduction-sum}, an example is given of a reduction (line 5) together with its reduction invariant (line 6) and post-state predicate (line 7). Intuitively, a reduction invariant is similar to a loop invariant. First, it must hold before the reduction starts. In our example this means that \halinline|count(x)| has the value \halinline|0|, which is ensured by the previous definition of \halinline|count| (line 4). Second, it must be preserved by each step of the reduction. In our example, \halinline|count| is bounded by the reduction variable. Finally, after each reduction variable has reached its maximum value, the reduction invariant should imply the post-state predicate of the function. For the example, note that the invariant implies the post-state predicate when \halinline|r| has reached the value \halinline|10|. The actual used value goes to \halinline|9|, and \halinline|r==10| indicates that the reduction is done.

\begin{lstlisting}[style=pvl, caption={The front-end \pvl code for the blur example (Listing~\ref{lst:blur-halide}). We omitted the \pvlinline|decreases| clauses for brevity.}, label=lst:front-end-blur]
pure int inp(int x, int y);
pure int inp_x_min(); pure int inp_x_max(); pure int inp_y_min(); pure int inp_y_max();
pure int blur_y_x_min(); pure int blur_y_x_max();
pure int blur_y_y_min(); pure int blur_y_y_max();

 ensures \result == (inp(x, y) + inp(x+1, y) + inp(x+2, y))/3;
pure int blur_x(int x, int y) = (inp(x, y) + inp(x+1, y) + inp(x+2, y))/3;

 ensures \result == ((inp(x, y) + inp(x+1, y) + inp(x+2, y))/3
   + (inp(x, y+1)   + inp(x+1, y+1)   + inp(x+2, y+1))/3
   + (inp(x,y+2) + inp(x+1,y+2) + inp(x+2,y+2))/3)/3;
pure int blur_y(int x, int y) = (blur_x(x, y) + blur_x(x, y+1) + blur_x(x, y+2))/3;

 requires inp_x_min() == blur_y_x_min() && inp_x_max() == blur_y_x_max()+2
   && inp_y_min() == blur_y_y_min() && inp_y_max() == blur_y_y_max()+2;
 ensures (\forall x, y; blur_y_x_min()<=x && x<blur_y_x_max() && blur_y_y_min()<=y && y<blur_y_y_max();
  blur_y(x,y) == ((inp(x, y) + inp(x+1, y) + inp(x+2, y))/3
   + (inp(x, y+1)   + inp(x+1, y+1)   + inp(x+2, y+1))/3
   + (inp(x, y+2) + inp(x+1, y+2) + inp(x+2, y+2))/3)/3);
void pipeline() { }
\end{lstlisting}

\subsection{Front-end verification approach}\label{sec:front-end}
For verifying the algorithm part of a \halide program, an annotated \halide algorithm is encoded into annotated \pvl code.
Listings~\ref{lst:front-end-blur} and \ref{lst:front-end-sum} show how \haliver translates the examples of Listings~\ref{lst:blur-halide} and \ref{lst:reduction-sum}, respectively. Input buffers are translated into abstract functions to verify the pipeline w.r.t.\ arbitrary input. The bounds of input buffers and functions are modelled via functions that are abstract if the bound is unknown or otherwise return a concrete value. For example, the \halinline|inp| buffer of the blur example is translated to a function \pvlinline|inp| in Listing \ref{lst:front-end-blur} (line 1), with its bounds 
represented by the pure functions on line 2.

Update-free \halide functions are translated directly into \pvlinline|pure| \pvl functions, and post-state predicates are translated into postconditions of these functions. In the example, \halinline|blur_x| and \halinline|blur_y| are translated to the functions on lines 6--7 and 9--12 of Listing~\ref{lst:front-end-blur}, respectively, and the \pvlinline|ensures| lines express the postconditions of those functions, using \pvlinline|\result| to refer to the expected result.

The pre- and postconditions of a \halide algorithm are translated into a \pvl lemma to be checked by \vercors. In the example, lines
14--19 of Listing~\ref{lst:front-end-blur} address the pre- and postconditions on lines 1--2 of Listing~\ref{lst:blur-halide}. On line 20, a method called \pvlinline|pipeline| is given, which represents the \halide pipeline.

For an update definition, references to itself are replaced by references to the previous definition, thus the output of one definition is the input of the next.

For a reduction, the initialisation and update parts are translated into separate functions, and reduction domain variables are explicitly added as function parameters. Listing~\ref{lst:front-end-sum} illustrates this for the \halinline|cnt| example. The function \pvlinline|count0| on line 8 corresponds to the initialisation (line 3 of Listing~\ref{lst:reduction-sum}), with the translated post-state predicate on line 6.
The function \pvlinline|count1r| (lines 13--14) corresponds to the update function (line 5 of Listing~\ref{lst:reduction-sum}). Note that the annotation on line 10 refers to the reduction domain. The reason for using references to \pvlinline|r-1| on line 14 is that the result of the whole computation corresponds to \pvlinline|r| with its maximum value \pvlinline|10| (see line 18). This is computed by recursively decrementing \pvlinline|r|. The invariant on line 6 of Listing~\ref{lst:reduction-sum}
is translated into the postcondition of \pvlinline|count1r| (line 11), reflecting that the invariant should hold after each reduction iteration. For the \pvlinline|decreases r| annotation added on line 12, \vercors will try to prove that this recursive function terminates. 
The reduction postcondition is represented by the \pvlinline|ensures|
annotation on line 16.

\remove{
In \vercors, \pvlinline|pure| functions can be used in annotations, and recursion-free \pvlinline|pure| functions do not need annotations themselves to be used. Therefore, the intermediate annotations are not needed by the front-end module, except for reduction invariants. However, the intermediate annotations are checked for validity.
}

\paragraph{\textbf{Guarantees}.}
For the front-end verification approach, \haliver straightforwardly encodes a \halide function without reductions, as it defines the function pointwise in \pvl. For reductions, \haliver mimics the iterative updates with recursion, as shown in the \halinline|cnt| example of Listings~\ref{lst:reduction-sum} and \ref{lst:front-end-sum}. \haliver adds \pvlinline|decreases| clauses to check that the recursive functions terminate.

With \haliver's approach, functional correctness of the algorithm part can be proven. Since memory safety depends on how a \halide algorithm is compiled into actual code according to a schedule, this is checked using the back-end verification approach.


\begin{lstlisting}[style=pvl, caption={The front-end \pvl code for the reduction example of Listing~\ref{lst:reduction-sum}.}, label=lst:front-end-sum]
  decreases;
pure int inp(int x, int y);
  decreases;
pure int inp_x_min(); pure int inp_x_max(); pure int inp_y_min(); pure int inp_y_max();

 ensures \result == 0;
 decreases;
pure int count0(int x) = 0;

 requires 0<=r && r<=10;
 ensures (0<=\result && \result<=r);
 decreases r;
pure int count1r(int x, int r) = r == 0 ? count0(x)
  : inp(x, r-1) > 0 ? count1r(x, r-1) + 1 : count1r(x, r-1);

 ensures (0<=\result && \result<=10);
 decreases;
pure int count(int x) = count1r(x, 10);
\end{lstlisting}

\subsection{Back-end verification approach}\label{sec:back-end}
\if@remove

\else
\begin{lstlisting}[style=halide, caption={\beginremove Example of permission and pre-state annotations that we automatically generate. The annotations on lines 3 and 9 should be provided by the user. The other annotations are added automatically.}, label=lst:back-end-generation]
f(x, y) = x + y;
f.context(Perm(f(x,y), 1\1); // Generated
f.ensures(f(x,y) == x + y));

f(x, 0) = f(x, x) + f(x, 0) + f(x, 1);
f.context(Perm(f(x,0), 1\1)); // Generated
f.context(\forall y; f.y.min<=y && y<f.y.max && y != 0; Perm(f(x,y), 1\2)); // Generated
f.requires(\forall y; f.y.min<=y && y<f.y.max; f(x,y) == x + y); // Generated
f.ensures(f(x,0) == 4*x + 1);

// The following annotation for f is used as context annotation wherever f is called:
 (\forall x, y; f.x.min<=y && y<f.x.min && f.y.min<=y && y<f.y.min;
   (y==0 ==> f(x,0) == 4*x + 1) && (y != 0 ==> f(x,y) == x + y));
\end{lstlisting}
\fi
For verifying a \halide algorithm with a schedule, \haliver adds annotations to the generated \C code that can be checked by \vercors. First, \haliver generates read and write permissions and preconditions for functions used in definitions. This generation of permissions makes it possible to keep the annotations of \halide algorithms concise, since the user does not have to specify permissions. Second, \haliver transforms the annotations and adds them to the intermediate representation used by the \halide compiler. Finally, \haliver adds the annotations to the code, during the code generation of the \halide compiler.

\begin{lstlisting}[style=halide, label=lst:back-end-loopnest, caption={A schedule for the blur example (Listing~\ref{lst:blur-halide}), together with the loop nest the \halide compiler produces, given in the intermediate representation of \halide. The \halinline|blur_y| bounds are assumed to be from 0 up to 1,024 for dimensions \halinline|x| and \halinline|y|.}]
blur_y.split(y, yo, yi, 8).parallel(yo).split(x, xo, xi, 2).unroll(xi);
blur_x.store_at(blur_y, yo).compute_at(blur_y, yi).split(x, xo, xi, 2).unroll(xi);
// Below is the loop nest produced (not part of the schedule)
produce blur_y:
  parallel y.yo in [0, 127]:
    store blur_x:
      for y.yi in [0, 7]:
        produce blur_x:
          for y:
            for x.xo in [0, 511]:
              unrolled x.xi in [0, 2]:
                blur_x(...) = ...
        consume blur_x:
          for x.xo in [0, 511]:
            unrolled x.xi in [0, 2]:
              blur_y(...) = ...
\end{lstlisting}
\paragraph{\textbf{Annotation generation.}}
Since \halide algorithms consist of pure point-wise functions, permissions are relatively straightforward: for a function \halinline|f(x,...)|,
\haliver generates the write permission
\halinline|Perm(f(x,...),1\1)|. For the blur example from Listing \ref{lst:blur-halide}, \haliver generates \halinline|blur_x.context(Perm(blur_x(x,y), 1\1)| and \halinline|blur_y.context(Perm(blur_y(x,y), 1\1)| for function \halinline|blur_x| and \halinline|blur_y|, respectively.

For update functions and reductions, \haliver generates (1) read permissions for function values that are not being updated, and (2) a pre-state predicate, using the post-state predicate of the previous update step. \remove{See for example Listing~\ref{lst:back-end-generation}. Lines 2, 6, 7 and 8 are generated automatically. Note that on line 7 we add \halinline|y!=0|, because otherwise we would have more than \halinline|1\1| write permission, which is not valid.}

Once a function is fully defined, read permission is given to all values wherever the function is used, along with a context predicate containing any intermediate annotations of the function. \remove{See for example, lines 12 and 13 of Listing~\ref{lst:back-end-generation}. However, since the intermediate representation of the \halide compiler strictly separates the creation of a function from when it is used, we choose to do this in the next step.}

\paragraph{\textbf{Transformation of annotations.}}
Next, \haliver transforms the annotations according to the schedule given by the user and associates them with the corresponding parts of the optimised \halide program expressed in \halide's intermediate language.

\haliver supports the \halinline|split|, \halinline|fuse|, \halinline|parallel|, \halinline|unroll|, \halinline|store_at|, \halinline|reorder| and \halinline|compute_at| scheduling directives. Of the most commonly used directives in the \halide example apps\footnote{\url{https://github.com/halide/Halide/tree/main/apps}}, only \halinline|vectorize| is not supported because \vercors does not yet support verification of vectorised code as produced by \halide.\footnote{The vectorize scheduling directive is the same as the unroll directive from the perspective of transforming annotations. So they can be treated exactly the same and already are in HaliVer. } With these directives, \haliver provides the means to verify optimised programs w.r.t.\ memory locality, parallelism and recomputation. This is the optimisation space in which \halide  resides~\cite{ragan-kelleyHalideDecouplingAlgorithms2017}. We illustrate the meaning of
these directives with an example. Listing~\ref{lst:back-end-loopnest} shows a schedule for blur on lines 1--2, and
below that the \textit{loop nest} structure of the resulting program. Loop nests are program statements of nested for loops. The loops can be sequentially executed or be parallelized, unrolled or vectorized. 
The allocation of space for a function result is indicated by \halinline|store|,
and \halinline|produce| and \halinline|consume| refer to writing and reading function results, respectively. This loop nesting corresponds to the actual code produced by the \halide compiler.

Assuming that the output dimensions in the example are both of size 1,024, the directive \halinline|split(y, yo, yi, 8)| (line 1 of Listing \ref{lst:back-end-loopnest}) splits the dimension \halinline|y| into two nested dimensions \halinline|y.yo| (line 5) and \halinline|y.yi| (line 7) of sizes \halinline|128| and \halinline|8|, respectively. \haliver similarly renames references to \halinline|y| in annotations. The \halinline|parallel(yo)| directive (line 1) expresses that \halinline|y.yo| should be executed in parallel (line 5). The \halinline|store_at(blur_y, yo)| directive (line 2) expresses that \halinline|blur_x| must be stored at the start of the \halinline|y.yo| loop (line 6). The directive \halinline|compute_at(blur_y, yi)| (line 2) defines that the values for \halinline|blur_x| should be produced at \halinline|y.yi| (line 8). The directive \halinline|unroll(xi)| (line 1 and 2) expresses that the dimension \halinline|xi| should be completely unrolled.

The \halinline|for| loops are sequential. In this example, \halinline|fuse| and \halinline|reorder| are not used; they express that two dimensions should be fused into one and the nesting order of the loops should be changed, respectively.

\haliver moves bottom-up through the program, constructing loop invariants for each loop by taking the constructed state predicates from the loop body and extending them with quantifications over the loop variables. Below, we give an example of this exact process for the blur example of Listing~\ref{lst:blur-halide}. Table~\ref{tbl:transformation} in the appendix explains the approach in a more general way.

\if@removetwo

\else
\begin{lstlisting}[style=pvl, float=tp, caption={The \C code and annotations the \halide compiler produces together with \haliver for the function \halinline|blur_y|, focussing on the \halinline|consume blur_x| node (see line 13 of Listing~\ref{lst:back-end-loopnest}). Listings~\ref{lst:blur-translation-back-end-1}--\ref{lst:blur-translation-back-end-3} from the appendix give the complete encoding for the \halinline|blur_y| pipeline.}, label=lst:blur-translation-consume-blur-x,escapeinside=``]
struct halide_dimension_t {int32_t min, max;};
struct buffer {int32_t dimensions;struct halide_dimension_t *dim;int32_t *host;};
int div_eucl(int x, int y);
//@ pure int hdiv(int x, int y) = y == 0 ? 0 : div_eucl(x, y);
//@ pure int p_i(int x);
/*@ ... // Buffers annotations
 context (\forall int x,int y;0<=x&&x<1026&&0<=y&&y<1026; inpb->host[y*1026+x]==p_i(y*1026+x));
 // Pipeline preconditions
 requires inpb->dim[0].min==blur_yb->dim[0].min&& inpb->dim[0].max==blur_yb->dim[0].max+2;
 requires inpb->dim[1].min==blur_yb->dim[1].min&& inpb->dim[1].max==blur_yb->dim[1].max+2;
 // Pipeline postconditions
 ensures (\forall int x, int y; 0<=x&&x<1024&&0<=y& y<1024; blur_yb->host[y*1024+x] == hdiv(
  hdiv(inpb->host[y*1026+x+1027]+inpb->host[y*1026+x+1028]+inpb->host[y*1026+x+1026],3)+ 
  hdiv(inpb->host[y*1026+x+2053]+inpb->host[y*1026+x+2054]+inpb->host[y*1026+x+2052],3)+ 
  hdiv(inpb->host[y*1026+x+1]+inpb->host[y*1026+x+2]+inpb->host[y*1026+x],3),3));@*/
int blur_3(struct buffer *inpb, struct buffer *blur_yb) {
 int32_t* blur_y = blur_yb->host;
 int32_t* inp = inpb->host;
 // produce blur_y
 #pragma omp parallel for
 for (int yo = 0; yo<0 + 128; yo++)
 ... // Annotations blur_y.y.yo
 {
   int64_t _2 = 10240;
   int32_t *blur_x = (int32_t  *)malloc(sizeof(int32_t )*_2);
   int32_t _t11 = (yo * 8);
   ... // Annotations blur_y.y.yi
   for (int yi = 0; yi<0 + 8; yi++)
   {... // produce blur_x
    // consume blur_x
    int32_t _t16 = (yi + _t11) * 512;
    int32_t _t15 = yi * 512;
 /*@ loop_invariant 0<=xo && xo<=0 + 512;
     loop_invariant (\forall* int x, int y; 0<=x && x<1024 && yo*8<=y && y<yo*8 + 10; 
      Perm(&blur_x[(y-yo*8)*1024+x], 1\2));
     loop_invariant (\forall int xo, int y; 0<=xo && xo<1024 && yo*8+yi<=y && y<=yo*8+yi+2;
      blur_x[(y-yo*8)*1024+xo] == hdiv(p_i(y*1026+xo) + p_i(y*1026+xo+1) + p_i(y*1026+xo+2),3));
     loop_invariant (\forall* int xif, int xof; 0<=xof && xof<512 && 0<=xif && xif<2;
      Perm(&blur_y[(yo*8+yi)*1024+xof*2+xif], 1\1));
     loop_invariant (\forall int xof, int xif; 0<=xof && xof<xo && 0<=xif && xif<2; blur_y[(yo*8+yi)*1024+xof*2+xif] == 
      hdiv(hdiv(p_i((yo*8+yi)*1026+xof*2+xif) + p_i((yo*8+yi)*1026+xof*2+xif+1) + p_i((yo*8+yi)*1026+xof*2+xif+2), 3) + 
      hdiv(p_i((yo*8+yi)*1026+xof*2+xif+1026) + p_i((yo*8+yi)*1026+xof*2+xif+1027) + p_i((yo*8+yi)*1026+xof*2+xif+1028), 3) + 
      hdiv(p_i((yo*8+yi)*1026+xof*2+xif+2052) + p_i((yo*8+yi)*1026+xof*2+xif+2053) + p_i((yo*8+yi)*1026+xof*2+xif+2054), 3), 3)); @*/ 
    for (int xo = 0; xo<0 + 512; xo++)
    {
     int32_t _t9 = (xo + _t15);
     blur_y[(xo + _t16) * 2] = div_eucl(blur_x[_t9 * 2] + blur_x[_t9 * 2 + 1024] + blur_x[_t9 * 2 + 2048], 3);
     blur_y[(xo + _t16) * 2 + 1] = div_eucl(blur_x[_t9 * 2 + 1] + blur_x[_t9 * 2 + 1025] + blur_x[_t9 * 2 + 2049], 3);
    } // for xo
   } // for yi
   free(blur_x);
 } // for yo
 return 0;
}
\end{lstlisting}
\fi

\paragraph{\textbf{Encoding of \halide program.}}
Finally, \haliver adds annotations to the \C code during the code generation of the \halide compiler. As an example, we show how \haliver adds annotations of the \halinline|blur_y| function of Listing~\ref{lst:blur-halide} with the schedule of Listing~\ref{lst:back-end-loopnest}.
The result of this can be found in Listing~\ref{lst:blur-translation-consume-blur-x}.
It shows the structure of the whole program, but is focussed on the code below the \halinline|consume blur_x| node (line 13 of Listing~\ref{lst:back-end-loopnest}). The complete \C code can be found \add{in the appendix }in Listings~\ref{lst:blur-translation-back-end-1}--\ref{lst:blur-translation-back-end-3}.

First, \haliver updates its pipeline annotations (lines 1--2 of Listing~\ref{lst:blur-halide}), to match the flattened array structure the \halide back-end uses, and adds them to the function contract (lines 8--15 of Listing~\ref{lst:blur-translation-consume-blur-x}). \haliver also uses the \halide definition of division (\pvlinline|hdiv|), i.e., Euclidean\footnote{The \halide compiler uses bit operators to define euclidean division. However, bit operators are not supported in \vercors, so \haliver uses an equivalent definition.}~\cite{leijenDivisionModulusComputer} with $x/0 \equiv 0$.

Next, \haliver transforms the annotations added to the \halinline|blur_y| function, before it adds them to any loop nest.
The \halide compiler flattens the two-dimensional function \halinline|blur_y(x,y)| into a one-dimensional array \pvlinline|blur_y[y*1024 + x]|, so \haliver
does the same for all function references in the annotations. Next, from the schedule, the directive \halinline|split(x, xo, xi, 2)| splits \halinline|x| into \halinline|xo| and \halinline|xi| of sizes \halinline|512| and \halinline|2|, respectively. A similar split is performed for \halinline|y|.  The generated annotation \halinline|context (Perm(blur_y(x,y), 1\1))| becomes \pvlinline|context Perm(&blur_y[(yo*8+yi) + xo*2 + xi], 1\1))|.

For the annotation \halinline|ensures(blur_y(x,y) == (((inp(x,y) + ...|, \haliver replaces the calls to \halinline|inp(x,y)| with calls to an abstract pure function \pvlinline|p_i|. This is done because quantification instantiation in \vercors can become unstable if \pvlinline|inp| is used frequently. Where \pvlinline|inp| is used in the code, \haliver adds annotations stating that \pvlinline|inp| and \pvlinline|p_i| have the same value (line 7 of Listing~\ref{lst:blur-translation-consume-blur-x}).

\haliver adds these annotations to the first loop nest, starting bottom up. In Listing~\ref{lst:back-end-loopnest}, this is \halinline|xi|, but since this loop is unrolled, additional annotations are not needed. After passing this loop nest, anything for \halinline|xi=0| and \halinline|xi=1| now holds. \haliver changes the annotations by quantifying over \halinline|xi|'s domain.  It uses \halinline|xif| as variable and changes any references to \halinline|xi| towards \halinline|xif|. The resulting permissions are \pvlinline|(\forall xif; 0<=xif && xif<2; Perm(blur_y[(yo*8+yi) + xo*2+xif], 1\1))|. The other annotations are processed in a similar way.

Next, \haliver arrives at the loop nest for \pvlinline|xo|, which needs loop invariants. First, the tool adds the bounds of the \pvlinline|xo| dimension (line 33 of Listing~\ref{lst:blur-translation-consume-blur-x}). The annotation is transformed depending on whether it was a \pvlinline|requires|, \pvlinline|ensures| or \pvlinline|context| annotation. The write permission (\pvlinline|context|), should hold before the loop starts and after the loop ends.
Therefore, \haliver adds the permission, but quantifies over dimension \pvlinline|xo|, which results in a loop invariant (lines 38--39 of Listing~\ref{lst:blur-translation-consume-blur-x}). The \pvlinline|ensure| annotation does not hold at the start of the loop, but after each iteration of the loop, one more value for \pvlinline|xo| holds. Therefore, \haliver quantifies over \pvlinline|xof| bounded by zero and the iteration variable \pvlinline|xo|, and replaces occurrences to \pvlinline|xo| with \pvlinline|xof|, which leads to a loop invariant (lines 40--43 of Listing~\ref{lst:blur-translation-consume-blur-x}). For loops above this loop nest, the \pvlinline|ensure| annotations hold for the whole domain of \pvlinline|xo|, resulting in \pvlinline|ensures (\forall xof, xif; 0<=xof && xof<512 && 0<=xif && xif<2; blur_y[(yo*8+yi)*1024+xof*2+xif] == ...|. This annotation is added to the parallel for loop (line 66 of Listing~\ref{lst:blur-translation-back-end-2}).

After constructing the \halinline|produce| node for \halinline|blur_y|, the \halinline|produce| node for \halinline|blur_x| is constructed in a similar way. The bound inferencer of \halide detects it only needs to calculate for \pvlinline|y| values of \pvlinline|8*y0+yi| up to \pvlinline|8*y0+yi+2|. The annotations are transformed respecting that fact. After the \halinline|produce| node, the \halinline|blur_x| is consumed (line 30 of Listing~\ref{lst:blur-translation-consume-blur-x}). So for each loop below the \halinline|consume| statement, \haliver adds read permission (lines 34--35 of Listing~\ref{lst:blur-translation-consume-blur-x}s) and the post-state predicate of \halinline|blur_x| (lines 36-37 of Listing~\ref{lst:blur-translation-consume-blur-x}) as context annotations. For the loop of \pvlinline|xo|, this means they are valid for any value of \pvlinline|xo|.

\paragraph{\textbf{Guarantees}.}
With the back-end verification approach, \haliver can prove that the optimised code produced by the \halide compiler is correct w.r.t.\ specifications. Memory safety is proven without any additional effort, as the permission annotations for this are generated automatically. For functional correctness, a specification needs to be provided. For any non-inlined function, an intermediate annotation is required to guide \vercors in correct functional verification.

The approach is sound, but not necessarily complete. One concern is that, since we have not formally proved the correctness of the transformation, our implementation could in principle be wrong. \haliver addresses this by keeping the \textit{pipeline} annotations very close to what the user has written as annotations. These pipeline annotations act as the formal contract that will be verified, and the user can inspect these at any time. If an intermediate annotation is not correctly transformed, the verification will fail, thus remaining sound but not complete. Of course we have not constructed any transformations to be wrong, but even if there is an oversight, we will remain sound.
Moreover, in Section~\ref{sec:experiments}, we show that our approach works for real world examples.

\begin{table}[t]
\centering
\caption{Number of lines of code and annotations for different \halide algorithms, schedules  and resulting programs, and the verification times required by \vercors to prove memory safety, given that no annotations are provided by the user.  The letters after each schedule denote the used directives: \texttt{\underline{c}ompute_at}, \texttt{\underline{f}use}, \texttt{\underline{p}arallel}, \texttt{\underline{r}eorder}, \texttt{\underline{s}plit}, \texttt{\underline{st}ore_at} and \texttt{\underline{u}nroll}. $F$ stands for verification failed. Times with $^{\dag}$ are inconsistent, i.e. they are succesfully verified, but can also sometimes fail or timeout. }
\label{tab:verification-results-memory} \small
 \centering
{
\tiny
\begin{tabular}{l l \vbar \vbar r \vbar r \vbar r r r r}
\hline \textbf{Name} & & \multicolumn{1}{l\vbar}{\textbf{\halide}} & \textbf{\textbf{Sched}}. & \multicolumn{3}{l}{\textbf{\C}} & \\
& & \textbf{LoC} & \textbf{Dir.} & \textbf{LoC} & \textbf{LoA.} & \textbf{Loops} & \textbf{T. (s).} \\ \hline \hline
blur & V0 & 38 & 0 & 178 &60 &2 & 18\\ \hline
 & V1-\{f,p\} & \ditto &2 &172 &56 &1 & 19\\ \hline
 & V2-\{c,p,r,s\} & \ditto &6 &212 &74 &6 & 29\\ \hline
 & V3-\{c,p,s,st,u\} & \ditto &8 &211 &72 &5 & 24\\ \hline
\hline
hist & V0 & 71 & 2 & 299 &98 &11 & 30\\ \hline
 & V1-\{c,p,r,u\} & \ditto &4 &308 &99 &11 & 38\\ \hline
 & V2-\{c,p,r,u\} & \ditto &6 &311 &105 &13 & 48\\ \hline
 & V3-\{c,p,r,u\} & \ditto &13 &312 &101 &13 & 48\\ \hline
\hline
conv\_ & V0 & 44 & 0 & 273 &148 &7 & 90\\ \hline
layer & V1-\{c,f,p,u\} & \ditto &4 &281 &145 &8 & 97\\ \hline
 & V2-\{p,r,s,u\} & \ditto &6 &302 &166 &10 & 209\\ \hline
 & V3-\{c,p,r,s,u\} & \ditto &15 &279 &148 &7 & 168\\ \hline
\hline
gemm & V0 & 70 & 0 & 218 &105 &3 & 41\\ \hline
 & V1-\{c,p,r,s\} & \ditto &8 &274 &136 &10 & 89\\ \hline
 & V2-\{c,p,r,s\} & \ditto &16 &342 &173 &19 & 196$^{\dag}$\\ \hline
 & V3-\{c,f,p,r,s,u\} & \ditto &24 &451 &221 &31 & F\\ \hline
\hline
auto\_ & V0 & 112 & 0 & 443 &118 &19 & 35\\ \hline
viz & V1-\{c\} & \ditto &9 &402 &139 &23 & 180\\ \hline
 & V2-\{c,p\} & \ditto &12 &440 &156 &27 & 170\\ \hline
 & V3-\{c,p,r,s\} & \ditto &27 &443 &152 &25 & 105\\ \hline
\hline
\multicolumn{2}{l \vbar \vbar}{camera\_pipe-\{c,p,r,s,st\} } & 345 & 27 & 701 &236 &25 & F\\ \hline
\hline
\multicolumn{2}{l \vbar \vbar}{bilateral\_grid-\{c,p,r,u\} } & 88 & 18 & 562 &180 &39 & 140\\ \hline
\hline
\multicolumn{2}{l \vbar \vbar}{depthwise\_separable\_conv-\{c,p,r,s\} } & 94 & 13 & 562 &315 &44 & 480\\ \hline
\hline
\end{tabular}
}
\end{table}

\begin{table}[t]
\centering
\caption{Number of lines of code and annotations for different \halide algorithms, schedules  and resulting programs, and the verification times required by \vercors.}
\label{tab:verification-results} \small
 \centering
{
\tiny
\begin{tabular}{l l \vbar \vbar r r \vbar r \vbar r \vbar r r r r \vbar \vbar r}
\hline Name & & \multicolumn{2}{l\vbar}{\halide} & \multicolumn{1}{l\vbar}{Front-end} & Sched. & \multicolumn{3}{l}{\C} & & LoA \\
& & LoC & LoA & T. (s) & LoC & LoC & LoA & Loops & T. (s) & incr. \\ \hline \hline
blur & V0 & 38 & 2 & 8 & 0 & 178 &63 &2 & 21 & 31.5x\\ \hline
 & V1-\{f,p\} & \ditto & \ditto & \ditto &2 &172 &58 &1 & 23 & 29.0x\\ \hline
 & V2-\{c,p,r,s\} & \ditto & \ditto & \ditto &6 &212 &83 &6 & 52 & 41.5x\\ \hline
 & V3-\{c,p,s,st,u\} & \ditto & \ditto & \ditto &8 &211 &79 &5 & 97$^{\dag}$ & 39.5x\\ \hline
\hline
hist & V0 & 71 & 10 & 8 & 0 & 299 &118 &11 & 34 & 11.8x\\ \hline
 & V1-\{c,p,r,u\} & \ditto & \ditto & \ditto &4 &308 &118 &11 & 47 & 11.8x\\ \hline
 & V2-\{c,p,r,u\} & \ditto & \ditto & \ditto &6 &311 &123 &13 & 56 & 12.3x\\ \hline
 & V3-\{c,p,r,u\} & \ditto & \ditto & \ditto &13 &312 &125 &13 & 64 & 12.5x\\ \hline
\hline
conv\_ & V0 & 44 & 7 & 8 & 0 & 273 &177 &7 & 111 & 25.3x\\ \hline
layer & V1-\{c,f,p,u\} & \ditto & \ditto & \ditto &4 &281 &174 &8 & 108 & 24.9x\\ \hline
 & V2-\{p,r,s,u\} & \ditto & \ditto & \ditto &6 &302 &204 &10 & 283 & 29.1x\\ \hline
 & V3-\{c,p,r,s,u\} & \ditto & \ditto & \ditto &15 &279 &177 &7 & 207 & 25.3x\\ \hline
\hline
gemm & V0 & 70 & 12 & 7 & 0 & 218 &120 &3 & 43 & 10.0x\\ \hline
 & V1-\{c,p,r,s\} & \ditto & \ditto & \ditto &8 &274 &169 &10 & 133 & 14.1x\\ \hline
 & V2-\{c,p,r,s\} & \ditto & \ditto & \ditto &16 &342 &230 &19 & 368 & 19.2x\\ \hline
 & V3-\{c,f,p,r,s,u\} & \ditto & \ditto & \ditto &24 &451 &310 &31 & F & 25.8x\\ \hline
\hline
auto\_ & V0 & 112 & 15 & 8 & 0 & 443 &158 &19 & 152$^{\dag}$ & 10.5x\\ \hline
viz & V1-\{c\} & \ditto & \ditto & \ditto &9 &402 &210 &23 & 216 & 14.0x\\ \hline
 & V2-\{c,p\} & \ditto & \ditto & \ditto &12 &440 &235 &27 & 230$^{\dag}$ & 15.7x\\ \hline
 & V3-\{c,p,r,s\} & \ditto & \ditto & \ditto &27 &443 &229 &25 & 192$^{\dag}$ & 15.3x\\ \hline
\hline
\end{tabular}

}
\end{table}

\section{Evaluation}\label{sec:experiments}
The goal of the evaluation of \haliver is four-fold. (1) We evaluate that the front-end verification approach of \haliver can verify functional correctness properties for a representative set of \halide algorithms. (2) For the back-end verification approach, the annotations that \haliver generates and transforms should lead to successful verification for a representative set of \halide programs, with schedules that use the most important scheduling directives in different combinations. (3) We evaluate the verification speed for front-end and back-end verification. (4) Lastly, we evaluate how many annotations are needed in \haliver compared to manually annotating the generated \C programs.\footnote{The experiments can be found at \url{https://github.com/sakehl/HaliVerExperiments}.}

\textbf{Set-up:} We used a machine with an 11th Gen Intel(R) Core(TM) i7-11800H @ 2.30GHz with 32GB running Ubuntu 23.04.

We used eight characteristic programs from the \halide repository.\footnote{\url{https://github.com/halide/Halide/tree/main/apps} \verb|gemm| is part of \verb|linear_algebra|.} These are representative \halide algorithm examples. They cover all scheduling directives supported by \haliver, in commonly-used combinations. We removed any scheduling directives that we do not support. 
As we discuss in Section~\ref{sec:triggers} of the appendix, \vercors is unable to deal with large dimensions that are unrolled, thus we removed some \halinline|unroll| directives as well.


The original schedule, as found in the \halide repository, is indicated with V3 if there are multiple schedules present. For five of these programs we defined annotations that express functional properties. These five programs are also evaluated with the standard schedule (V0), which tries to inline functions as much as possible, and two additional schedules (V1 and V2) we constructed.

\textbf{Memory safety results:}
We evaluate 8 \halide programs, with in total 23 schedules, and prove data race freedom and memory safety for 21 of them. No user provided annotations are needed. The results can be found in Table~\ref{tab:verification-results-memory}.

For each case, we provide: the number of lines of code (LoC)\footnote{These lines are counted automatically and indicate the size of the programs.} for the \halide algorithm, without the schedule and number of scheduling directives (Sched.\ Dir.). For the generated programs (\C) we list: lines of code (LoC), lines of annotations (LoA.), number of (parallel) loops (Loops). These numbers indicate how large programs tend to become w.r.t.\ \halide algorithms, and how much annotation effort would be required to manually annotate the programs. Verification running times (T.\ (s)) are given in seconds, averaged over five runs.

For \texttt{camera\_pipe}, \vercors gives a verification failure. It could not prove a \pvlinline|loop_invariant|, but after simplifying parts of the generated \C program not related to this specific invariant, it leads to a successful verification. This indicates that
the program is too complex for the underlying solvers. We also coded this example in similar \pvl code instead of \C, which verifies in 193s. We suspect the failure is caused by quantifier instantiation, which instantiates too many quantifiers, resulting in the SMT solver on which \vercors relies stopping the exploration of quantifiers that are needed for successful verification.

For \texttt{gemm V3}, verification fails due to \vercors not sufficiently rewriting annotations of the \halinline|fuse| directive. This is further explained in section \ref{sec:triggers}.

\textbf{Functional correctness results:}
Next, we evaluate five\footnote{The other three algorithms from the memory safety results are typical image processing pipelines. They are therefore less suitable for checking functional correctness and are not used here.} algorithms with annotations and 20 schedules, both for the front-end and back-end. \haliver proves functional correctness for the front-end, and both functional correctness and data race freedom and memory safety for the back-end for 19 of the 20 schedules. These results are given in Table~\ref{tab:verification-results}. The table additionally has the amount of user provided annotations (LoA.) and the last column (Ann. incr.) indicates the growth of the annotations. The annotations of the C file (LoC) contain both the generated annotations, which are already present in Table~\ref{tab:verification-results-memory} and the transformed user annotations.
For optimised programs, the annotation size is strongly related to the number of loops, as each loop needs its own loop invariants.
Front-end verification is successful for all examples and is relatively fast compared to back-end verification.
In verification of the \C files produced by the back-end verification approach, time increases as the number of scheduling directives increases.  Here, \texttt{gemm V3} also fails for the same reason
as outlined above.

\textbf{Inconsistent results:}
For \texttt{gemm V2} for the memory benchmarks and for \texttt{blur V3} and  \texttt{auto\_viz V0, V2} and \texttt{V3}, 
\vercors does not always succeed with the verification. In the case of \texttt{gemm V2}, the verification sometimes hangs, which is timed out after 10 minutes. In the other cases, \vercors sometimes gave a verification failure. This inconsistency is due to the non-deterministic nature of the underlying SMT solvers.

\textbf{Conclusions:}
With the front-end verification approach of \haliver we are able to prove functional correctness properties for representative \halide algorithms. Using \haliver's back-end verification approach, the tool provides correct annotations for the generated \C programs. \vercors successfully verifies all but two programs. However, in the unsuccessful cases, \haliver runs into limitations of the underlying tools. The verified programs are all verified within ten minutes. Finally, the manual annotation effort required is an order of magnitude larger than the effort required for \haliver's approach.

\section{Related Work}\label{sec:related-work}

%

There is much work on optimising program transformations, either applied automatically or manually~\cite{bacon94,Kowa03},
sometimes using scheduling languages~\cite{ragan-kelleyHalideDecouplingAlgorithms2017,ragan-kelleyHalideLanguageCompiler2013,chenTVMAutomatedEndtoEnd2018,tiramisu,chill,graphit,hagedornFireironDataMovementAwareScheduling2020}.
The vast majority of this does not address functional correctness.

Work on functional correctness consists of techniques that apply verification every time a program is transformed, and techniques that
verify the compiler.

Liu \emph{et al.}~\cite{liuVerifiedTensorprogramOptimization2022} propose an approach inspired by scheduling languages, with proof obligations generated when a program is optimised, for automatic verification using \coq. The \cogent language~\cite{cogent} uses refinement proofs, to be verified in \isabellehol. However, it does not separate algorithms from schedules.
In~\cite{namjoshi2016loopy,namjoshi2021self} an integer constraint solver and a proof checker are used, respectively, to verify the transformation of a program. In all these approaches, semantics-preservation is the focus, as opposed to specifying the intended behaviour. Model-to-model transformations can be verified w.r.t.\ the preservation of functional properties~\cite{de2016verifying}. However, that work targets models, not code.

Regarding the verification of compilers, \compcert~\cite{leroy2009formally} is a framework involving a formally verified C compiler. In~\cite{newcombVerifyingImprovingHalide2020}, \halide's Term Rewriting System, used to reason about the applicability of schedules, is verified using Z3 and Coq. These approaches do not require verification every time an optimisation is applied, but verifying the compiler is time-consuming and complex, and has to be redone whenever the compiler is updated. Furthermore, they focus on semantics-preservation,
not the intended behaviour of individual programs.

\alpinist~\cite{sakarAlpinistAnnotationAwareGPU2022} is most closely related. This tool automatically optimises \pvl code, along with its annotations, for verification with \vercors. It allows the specification of intended behaviour, but it does not separate algorithms from schedules, forcing the user to reason about the technical details of parallelisation.


%
%
%



\section{Conclusions \& Future Work}\label{sec:Discussion}
We presented \haliver, a tool for verifying optimised code by exploiting the strengths of scheduling languages and deductive verification. It allows focussing on functionality when annotating programs, keeping annotations succinct.

For future work, we want to extend the \haliver tool with aspects not directly supported by \vercors, such as vectorisation. The master thesis of~\cite{reinkingFormalSemanticsHalide2020} defines a natural semantics for \halide. We want to formalise our front-end \pvl encoding with an axiomatic semantics to match this semantics.  We also want to investigate the inconsistent results 
and see whether annotations with quantifiers can be rephrased to allow \vercors to be more consistent. In this work we have focussed on parallel CPU code, but we have designed our approach to be extendable to GPU code produced by \halide.

With the current expressiveness of the annotations, when reduction domains are present, \haliver proves functional correctness for specific inputs. 
For example, in Listing~\ref{lst:reduction-sum} we can prove that \halinline|count(x) == 9| if we require that \halinline|input(x,y) == x|.
This can also be done for any input if the reduction domain is of known size, but then many annotations are needed. To make the annotations concise, a user needs to be able to use axiomatic data types\footnote{\url{https://vercors.ewi.utwente.nl/wiki/\#axiomatic-data-types}} and pure functions in their annotations. We expect that these annotations can be similarly transformed by our approach, and that is thus orthogonal to this contribution, but this is planned as future work.

Most \halide programs use floating point numbers. These are currently modelled as reals in \vercors. How to efficiently verify programs with floats using deductive verifiers is still an open research question. Once this is addressed, \haliver will be able to give better guarantees.

We require that the bounds of a \halide program are set to concrete values for our back-end verification approach. \haliver transforms the annotations the same way for not know bounds, but the underlying tools have difficulty verifying these programs. With unknown bounds, we end up with nonlinear arithmetic due to the flattening of multi-dimensional functions on one-dimensional arrays. This is generally undecidable, so the SMT solvers that \vercors rely on cannot handle it. We will investigate if there are ways to tackle this in our domain-specific case.



\subsubsection{Acknowledgements}
This work is carried out in the context of the NWO TTW ChEOPS project 17249.
We want to thank Jan Martens for their discussions and feedback on this work.

 \bibliographystyle{splncs04}
 \bibliography{bibliography}

\newpage
\appendix
\section{Quantification \& Triggers}\label{sec:triggers}
The verification approach of \haliver relies heavily on modelling the specification with \pvlinline|forall| quantifiers. 
For a quantifier to be realised, a \textit{trigger}~\cite{dross2012reasoning} must be instantiated, which determines when exactly \vercors can use the quantifier. For example, \pvlinline|\forall i; 0<=i<10; a[i] >= 0| gets the trigger \pvlinline|a[i]| and when \vercors sees the value \pvlinline|a[5]| it can test the quantifier for \pvlinline|i=5| to see if it is valid. A condition like \pvlinline|\forall i, j; 0<=i<10 && 0<=j<5; a[5*i+j]| has no trigger, because when \vercors sees \pvlinline|a[5]| it cannot know which values \pvlinline|i| and \pvlinline|j| should have. This can be overcome in this case because the quantifier is equivalent to \pvlinline|\forall i_j; 0<=i_j<10*5); a[i_j]|, for which \pvlinline|a[i_j]| can be used as a trigger. \vercors rewrites the conditions for some of these cases. Also, since \vercors introduces more quantifiers when encoding parallel blocks, this rewriting \textit{must} be done by \vercors.

In some cases, quantifiers are introduced for which no rewriting rules exist yet. For example, when the \halinline|fuse| directive is used, it can introduce expressions such as \pvlinline|\forall i, j; 0<=i<10 && 3<=j<6; a[5*i+(j 

Additionally, with trigger instantiation we may run into the problem of matching loops \cite{becker2019axiom}. For example, the quantifier \pvlinline|\forall i;...f[i] == x[i] + y[i+1]|, we could use \pvlinline|f| as a trigger, but if a quantifier with \pvlinline|\forall i; ... x[i] == f[i+1]| has a trigger \pvlinline|x[i]|, the first instantiated trigger could trigger the second, which loops back to the first trigger and continues indefinitely. We believe that we run into this problem when verifying \halide files that unroll a large dimension with the \pvlinline|unroll| directive. For example, if we want to prove that \pvlinline| i; 0<=x && x<10; f[x] == x| and we have unrolled the function f like \pvlinline|f[0] = 0; ... f[9] = 9|, the verifier needs to realise that these 10 lines of code \textit{together} help prove the quantifier. In future work we intend to investigate how to deal with this problem. For now, we have chosen to do our evaluation in section~\ref{sec:experiments} only for small unrolled dimensions, as \vercors gives unreliable results otherwise.
\section{Code Examples}
Listings~\ref{lst:blur-translation-back-end-1}--\ref{lst:blur-translation-back-end-3} present an example of code produced by \haliver. This code is a translation of Listing~\ref{lst:blur-halide}, together with the schedule as given in Listing~\ref{lst:back-end-loopnest}.

\if@removetwo
\begin{lstlisting}[style=pvl, float=tp, caption={The \C code and annotations the \halide compiler produces together with \haliver for the function \halinline|blur_y|, focussing on the \halinline|consume blur_x| node (see line 13 of Listing~\ref{lst:back-end-loopnest}). Listings~\ref{lst:blur-translation-back-end-1}--\ref{lst:blur-translation-back-end-3} from the appendix give the complete encoding for the \halinline|blur_y| pipeline.}, label=lst:blur-translation-consume-blur-x,escapeinside=``]
struct halide_dimension_t {int32_t min, max;};
struct buffer {int32_t dimensions;struct halide_dimension_t *dim;int32_t *host;};
int div_eucl(int x, int y);
//@ pure int hdiv(int x, int y) = y == 0 ? 0 : div_eucl(x, y);
//@ pure int p_i(int x);
/*@ ... // Buffers annotations
 context (\forall int x,int y;0<=x&&x<1026&&0<=y&&y<1026; inpb->host[y*1026+x]==p_i(y*1026+x));
 // Pipeline preconditions
 requires inpb->dim[0].min==blur_yb->dim[0].min&& inpb->dim[0].max==blur_yb->dim[0].max+2;
 requires inpb->dim[1].min==blur_yb->dim[1].min&& inpb->dim[1].max==blur_yb->dim[1].max+2;
 // Pipeline postconditions
 ensures (\forall int x, int y; 0<=x&&x<1024&&0<=y& y<1024; blur_yb->host[y*1024+x] == hdiv(
  hdiv(inpb->host[y*1026+x+1027]+inpb->host[y*1026+x+1028]+inpb->host[y*1026+x+1026],3)+ 
  hdiv(inpb->host[y*1026+x+2053]+inpb->host[y*1026+x+2054]+inpb->host[y*1026+x+2052],3)+ 
  hdiv(inpb->host[y*1026+x+1]+inpb->host[y*1026+x+2]+inpb->host[y*1026+x],3),3));@*/
int blur_3(struct buffer *inpb, struct buffer *blur_yb) {
 int32_t* blur_y = blur_yb->host;
 int32_t* inp = inpb->host;
 // produce blur_y
 #pragma omp parallel for
 for (int yo = 0; yo<0 + 128; yo++)
 ... // Annotations blur_y.y.yo
 {
   int64_t _2 = 10240;
   int32_t *blur_x = (int32_t  *)malloc(sizeof(int32_t )*_2);
   int32_t _t11 = (yo * 8);
   ... // Annotations blur_y.y.yi
   for (int yi = 0; yi<0 + 8; yi++)
   {... // produce blur_x
    // consume blur_x
    int32_t _t16 = (yi + _t11) * 512;
    int32_t _t15 = yi * 512;
 /*@ loop_invariant 0<=xo && xo<=0 + 512;
     loop_invariant (\forall* int x, int y; 0<=x && x<1024 && yo*8<=y && y<yo*8 + 10; 
      Perm(&blur_x[(y-yo*8)*1024+x], 1\2));
     loop_invariant (\forall int xo, int y; 0<=xo && xo<1024 && yo*8+yi<=y && y<=yo*8+yi+2;
      blur_x[(y-yo*8)*1024+xo] == hdiv(p_i(y*1026+xo) + p_i(y*1026+xo+1) + p_i(y*1026+xo+2),3));
     loop_invariant (\forall* int xif, int xof; 0<=xof && xof<512 && 0<=xif && xif<2;
      Perm(&blur_y[(yo*8+yi)*1024+xof*2+xif], 1\1));
     loop_invariant (\forall int xof, int xif; 0<=xof && xof<xo && 0<=xif && xif<2; blur_y[(yo*8+yi)*1024+xof*2+xif] == 
      hdiv(hdiv(p_i((yo*8+yi)*1026+xof*2+xif) + p_i((yo*8+yi)*1026+xof*2+xif+1) + p_i((yo*8+yi)*1026+xof*2+xif+2), 3) + 
      hdiv(p_i((yo*8+yi)*1026+xof*2+xif+1026) + p_i((yo*8+yi)*1026+xof*2+xif+1027) + p_i((yo*8+yi)*1026+xof*2+xif+1028), 3) + 
      hdiv(p_i((yo*8+yi)*1026+xof*2+xif+2052) + p_i((yo*8+yi)*1026+xof*2+xif+2053) + p_i((yo*8+yi)*1026+xof*2+xif+2054), 3), 3)); @*/ 
    for (int xo = 0; xo<0 + 512; xo++)
    {
     int32_t _t9 = (xo + _t15);
     blur_y[(xo + _t16) * 2] = div_eucl(blur_x[_t9 * 2] + blur_x[_t9 * 2 + 1024] + blur_x[_t9 * 2 + 2048], 3);
     blur_y[(xo + _t16) * 2 + 1] = div_eucl(blur_x[_t9 * 2 + 1] + blur_x[_t9 * 2 + 1025] + blur_x[_t9 * 2 + 2049], 3);
    } // for xo
   } // for yi
   free(blur_x);
 } // for yo
 return 0;
}
\end{lstlisting}
\else

\fi

\begin{lstlisting}[style=pvl, float=ht, caption={Back-end \C code with annotations provided by \haliver for the blur example of Listing~\ref{lst:blur-halide} (1/3).}, label=lst:blur-translation-back-end-1,escapeinside=``,firstnumber=auto]
#include <stdint.h>
#include <stdlib.h>


// Euclidean division is defined internally in VerCors
//@ pure int hdiv(int x, int y) = y == 0 ? 0 : \euclidean_div(x, y);
/*@
  requires y != 0;
  ensures \result == \euclidean_div(x, y);
@*/
inline int div_eucl(int x, int y)
{
    int q = x/y;
    int r = x%y;
    return r<0 ? q + (y > 0 ? -1 : 1) : q;
}

struct halide_dimension_t {int32_t min, max;};
struct buffer {int32_t dimensions;struct halide_dimension_t *dim;int32_t *host;};
 pure int p_i(int x);
/*@
 // Buffer Annotations
 context inpb != NULL ** Perm(inpb, 1\2);
 context Perm(inpb->dim, 1\2) ** inpb->dim != NULL;
 context \pointer_length(inpb->dim) == 2;
 context Perm(inpb->host, 1\2) ** inpb->host != NULL;
 context Perm(&inpb->dim[0], 1\2);
 context Perm(inpb->dim[0].min, 1\2) ** Perm(inpb->dim[0].max, 1\2);
 context Perm(&inpb->dim[1], 1\2);
 context Perm(inpb->dim[1].min, 1\2) ** Perm(inpb->dim[1].max, 1\2);
 context \pointer_length(inpb->host) == 1026*1026;
 context blur_yb != NULL ** Perm(blur_yb, 1\2);
 context Perm(blur_yb->dim, 1\2) ** blur_yb->dim != NULL;
 context \pointer_length(blur_yb->dim) == 2;
 context Perm(blur_yb->host, 1\2) ** blur_yb->host != NULL;
 context Perm(&blur_yb->dim[0], 1\2);
 context Perm(blur_yb->dim[0].min, 1\2) ** Perm(blur_yb->dim[0].max, 1\2);
 context Perm(&blur_yb->dim[1], 1\2);
 context Perm(blur_yb->dim[1].min, 1\2) ** Perm(blur_yb->dim[1].max, 1\2);
 context \pointer_length(blur_yb->host) == 1024*1024;
 context blur_yb->host != inpb->host;
 context inpb->dim[0].min == 0 && inpb->dim[0].max == 1026;
 context inpb->dim[1].min == 0 && inpb->dim[1].max == 1026;
 context (\forall* int x, int y; 0<=x && x<1026 && 0<=y && y<1026; Perm(&inpb->host[y*1026 + x], 1\2));
 context (\forall int x, int y; 0<=x && x<1026 && 0<=y && y<1026; inpb->host[y*1026 + x] == p_i(y*1026 + x));
 context blur_yb->dim[0].min == 0 && blur_yb->dim[0].max == 1024;
 context blur_yb->dim[1].min == 0 && blur_yb->dim[1].max == 1024;
 context (\forall* int x, int y; 0<=x && x<1024 && 0<=y && y<1024; Perm(&blur_yb->host[y*1024 + x], 1\1));
 // Pipeline preconditions
 requires inpb->dim[0].min == blur_yb->dim[0].min  && inpb->dim[0].max == blur_yb->dim[0].max+2;
  requires inpb->dim[1].min == blur_yb->dim[1].min  && inpb->dim[1].max == blur_yb->dim[1].max+2;
 // Pipeline postconditions
 ensures (\forall int x, int y; 0<=x && x<1024 && 0<=y && y<1024; blur_yb->host[y*1024 + x] == hdiv(hdiv(inpb->host[y*1026 + x + 1027] + inpb->host[y*1026 + x + 1028] + inpb->host[y*1026 + x + 1026], 3) + (hdiv(inpb->host[y*1026 + x + 2053] + inpb->host[y*1026 + x + 2054] + inpb->host[y*1026 + x + 2052], 3) + hdiv(inpb->host[y*1026 + x + 1] + inpb->host[y*1026 + x + 2] + inpb->host[y*1026 + x], 3)), 3));
@*/
int blur_3(struct buffer *inpb, struct buffer *blur_yb) {
\end{lstlisting}

\begin{lstlisting}[style=pvl, float=ht, caption={Back-end \C code with annotations provided by \haliver for the blur example of Listing~\ref{lst:blur-halide} (2/3).}, label=lst:blur-translation-back-end-2,escapeinside=``,firstnumber=last]
 int32_t* _blur_y = blur_yb->host;
 int32_t* _inp = inpb->host;
 // produce blur_y
 #pragma omp parallel for
 for (int yo = 0; yo<0 + 128; yo++)
 /*@
  context 0<=yo && yo<0 + 128;
  context (\forall* int x, int y; 0<=x && x<1026 && 0<=y && y<1026; Perm(&_inp[y*1026 + x], 1\(2*128)));
  context (\forall int x, int y; 0<=x && x<1026 && 0<=y && y<1026; _inp[y*1026 + x] == p_i(y*1026 + x));
  context (\forall* int xif, int xof, int yif; ((((0<=yif && yif<8) && 0<=xof) && xof<512) && 0<=xif) && xif<2; Perm(&_blur_y[((yo*8 + yif)*1024) + xof*2 + xif], 1\1));
  ensures (\forall int yif, int xof, int xif; ((((0<=yif && yif<8) && 0<=xof) && xof<512) && 0<=xif) && xif<2; _blur_y[((yo*8 + yif)*1024) + xof*2 + xif] == hdiv(hdiv(p_i((yo*8 + yif)*1026 + xof*2 + xif) + (p_i((yo*8 + yif)*1026 + xof*2 + xif + 1) + p_i((yo*8 + yif)*1026 + xof*2 + xif + 2)), 3) + (hdiv(p_i((yo*8 + yif)*1026 + xof*2 + xif + 1026) + (p_i((yo*8 + yif)*1026 + xof*2 + xif + 1027) + p_i((yo*8 + yif)*1026 + xof*2 + xif + 1028)), 3) + hdiv(p_i((yo*8 + yif)*1026 + xof*2 + xif + 2052) + (p_i((yo*8 + yif)*1026 + xof*2 + xif + 2053) + p_i((yo*8 + yif)*1026 + xof*2 + xif + 2054)), 3)), 3));
 @*/
 {
  {
   int64_t _2 = 10240;
   int32_t *_blur_x = (int32_t  *)malloc(sizeof(int32_t )*_2);
   int32_t _t11 = (yo * 8);
   /*@
    loop_invariant 0<=yi && yi<=0 + 8;
    loop_invariant (\forall* int x, int y; 0<=x && x<1026 && 0<=y && y<1026; Perm(&_inp[y*1026 + x], 1\(2*128)));
    loop_invariant (\forall int x, int y; 0<=x && x<1026 && 0<=y && y<1026; _inp[y*1026 + x] == p_i(y*1026 + x));
    loop_invariant (\forall* int x, int y; 0<=x && x<1024 && yo*8<=y && y<yo*8 + 10; Perm(&_blur_x[((y - yo*8)*1024) + x], 1\1));
    loop_invariant (\forall* int xif, int xof, int yif; 0<=yif && yif<8 && 0<=xof && xof<512 && 0<=xif && xif<2; Perm(&_blur_y[((yo*8 + yif)*1024) + xof*2 + xif], 1\1));
    loop_invariant (\forall int yif, int xof, int xif; 0<=yif && yif<yi && 0<=xof && xof<512 && 0<=xif && xif<2; _blur_y[((yo*8 + yif)*1024) + xof*2 + xif] == hdiv(hdiv(p_i((yo*8 + yif)*1026 + xof*2 + xif) + (p_i((yo*8 + yif)*1026 + xof*2 + xif + 1) + p_i((yo*8 + yif)*1026 + xof*2 + xif + 2)), 3) + (hdiv(p_i((yo*8 + yif)*1026 + xof*2 + xif + 1026) + p_i((yo*8 + yif)*1026 + xof*2 + xif + 1027) + p_i((yo*8 + yif)*1026 + xof*2 + xif + 1028), 3) + hdiv(p_i((yo*8 + yif)*1026 + xof*2 + xif + 2052) + p_i((yo*8 + yif)*1026 + xof*2 + xif + 2053) + p_i((yo*8 + yif)*1026 + xof*2 + xif + 2054), 3)), 3));
   @*/
   for (int yi = 0; yi<0 + 8; yi++)
   {
    // produce blur_x
    int32_t _t12 = (yi + _t11);
    /*@
     loop_invariant _t12<=y && y<=_t12 + 3;
     loop_invariant (\forall* int x, int y; 0<=x && x<1026 && 0<=y && y<1026; Perm(&_inp[y*1026 + x], 1\(2*128)));
     loop_invariant (\forall int x, int y; 0<=x && x<1026 && 0<=y && y<1026; _inp[y*1026 + x] == p_i(y*1026 + x));
     loop_invariant (\forall* int xif, int xof, int yf; yo*8 + yi<=yf && yf<yo*8 + yi + 3 && 0<=xof && xof<512 && 0<=xif && xif<2; Perm(&_blur_x[(yf - yo*8)*1024 + xof*2 + xif], 1\1));
     loop_invariant (\forall int yf, int xof, int xif; yo*8 + yi<=yf && yf<y && 0<=xof && xof<512 && 0<=xif && xif<2; _blur_x[(yf - yo*8)*1024 + xof*2 + xif] == hdiv(p_i((yf*513 + xof)*2 + xif) + p_i((yf*513 + xof)*2 + xif + 1) + p_i((yf*513 + xof)*2 + xif + 2), 3));
    @*/ 
    for (int y = _t12; y<_t12 + 3; y++)
    {
\end{lstlisting}
\begin{lstlisting}[style=pvl, float=ht, caption={Back-end \C code with annotations provided by \haliver for the blur example of Listing~\ref{lst:blur-halide} (3/3).}, label=lst:blur-translation-back-end-3,escapeinside=``,firstnumber=last]
     int32_t _t14 = ((y - _t11) * 512);
     int32_t _t13 = (y * 513);
     /*@
      loop_invariant 0<=xo && xo<=0 + 512;
      loop_invariant (\forall* int x, int y; 0<=x && x<1026 && 0<=y && y<1026; Perm(&_inp[y*1026 + x], 1\(2*128)));
      loop_invariant (\forall int x, int y; 0<=x && x<1026 && 0<=y && y<1026; _inp[y*1026 + x] == p_i(y*1026 + x));
      loop_invariant (\forall* int xif, int xof; 0<=xof && xof<512 && 0<=xif && xif<2; Perm(&_blur_x[(y - yo*8)*1024 + xof*2 + xif], 1\1));
      loop_invariant (\forall int xof, int xif; 0<=xof && xof<xo && 0<=xif && xif<2; _blur_x[(y - yo*8)*1024 + xof*2 + xif] == hdiv(p_i(y*1026 + xof*2 + xif) + p_i(y*1026 + xof*2 + xif + 1) + p_i(y*1026 + xof*2 + xif + 2), 3));
     @*/ 
     for (int xo = 0; xo<0 + 512; xo++)
     {
      int32_t _t7 = xo + _t13;
      _blur_x[((xo + _t14) * 2)] = div_eucl(_inp[_t7 * 2] + _inp[_t7 * 2 + 1] + _inp[_t7 * 2 + 2], 3);
      _blur_x[(((xo + _t14) * 2) + 1)] = div_eucl(_inp[_t7 * 2 + 1] + _inp[_t7 * 2 + 2] + _inp[_t7 * 2 + 3], 3);
     } // for xo
    } // for y
    // consume blur_x
    int32_t _t16 = (yi + _t11) * 512;
    int32_t _t15 = yi * 512;
    /*@
     loop_invariant 0<=xo && xo<=0 + 512;
     loop_invariant (\forall* int x, int y; 0<=x && x<1024 && yo*8<=y && y<yo*8 + 10; Perm(&_blur_x[(y - yo*8)*1024 + x], 1\2));
     loop_invariant (\forall int xo, int y; 0<=xo && xo<1024 && yo*8 + yi<=y && y<=yo*8 + yi + 2; _blur_x[(y - yo*8)*1024 + xo] == hdiv(p_i(y*1026 + xo) + p_i(y*1026 + xo + 1) + p_i(y*1026 + xo + 2), 3));
     loop_invariant (\forall* int xif, int xof; 0<=xof && xof<512 && 0<=xif && xif<2; Perm(&_blur_y[(yo*8 + yi)*1024 + xof*2 + xif], 1\1));
     loop_invariant (\forall int xof, int xif; 0<=xof && xof<xo && 0<=xif && xif<2; _blur_y[(yo*8 + yi)*1024 + xof*2 + xif] == hdiv(hdiv(p_i((yo*8 + yi)*1026 + xof*2 + xif) + p_i((yo*8 + yi)*1026 + xof*2 + xif + 1) + p_i((yo*8 + yi)*1026 + xof*2 + xif + 2), 3) + hdiv(p_i((yo*8 + yi)*1026 + xof*2 + xif + 1026) + p_i((yo*8 + yi)*1026 + xof*2 + xif + 1027) + p_i((yo*8 + yi)*1026 + xof*2 + xif + 1028), 3) + hdiv(p_i((yo*8 + yi)*1026 + xof*2 + xif + 2052) + p_i((yo*8 + yi)*1026 + xof*2 + xif + 2053) + p_i((yo*8 + yi)*1026 + xof*2 + xif + 2054), 3), 3));
    *@/
    for (int xo = 0; xo<0 + 512; xo++)
    {
     int32_t _t9 = (xo + _t15);
     _blur_y[(xo + _t16) * 2] = div_eucl(_blur_x[_t9 * 2] + _blur_x[_t9 * 2 + 1024] + _blur_x[_t9 * 2 + 2048], 3);
     _blur_y[(xo + _t16) * 2 + 1] = div_eucl(_blur_x[_t9 * 2 + 1] + _blur_x[_t9 * 2 + 1025] + _blur_x[_t9 * 2 + 2049], 3);
    } // for xo
   } // for yi
   free(_blur_x);
  } // alloc _blur_x
 } // for yo
 return 0;
}
\end{lstlisting}

\section{Transformation of annotations for back-end verification approach}
In Table~\ref{tbl:transformation} we explain how the annotations are transformed in a general sense for back-end verification.

\begin{table}
\caption{Here we show how to transform a function definition's annotations. When we write \halinline|E =>> E'| we mean that we replace the expression (or annotation) \halinline|E| with the expression \halinline|E'|. By \halinline|E[x =>> e]| we mean that we replace every occurrence of the variable \halinline|x| in \halinline|E| with \halinline|e|. If a dimension \halinline|x| is not scheduled, it is marked as \halinline|serial(x)|, which represents a sequential loop.\label{tbl:transformation}}
{
\scriptsize
\begin{tabular}{ p{1.2in} \vbar p{3.5in} }

\textbf{Loop nests} & A loop nest is created for each function definition. We store the annotations which were attached to the function definition. We traverse the loop nest from bottom to top, transforming the annotations at each loop. The loops in the loop nest that we have not yet visited are called the outer loops. We always do two things when we pass through a loop: 1. append annotations based on the current stored annotations; 2. modify the stored annotations for use in the outer loops. \\ \hline \hline
\textbf{Without reductions} & When passing a loop for the dimension \halinline|x|, \haliver transforms the annotation \halinline|P| for the remaining outer loops by quantifying over \halinline|x|: \halinline|P =>> (\forall xf;x.min<=xf<x.max; P[x =>> xf])| \\ \hline
\halinline|parallel(x)| & The parallel block contract uses the stored annotations as is. \\ \hline
\halinline|unroll(x)| & The code is unrolled and requires no further annotations \\ \hline
\halinline|serial(x)|
   & The annotation \halinline|P| becomes loop invariant and we quantify over |x|. The limits depend on whether it is a pre- or postcondition. \\
\halinline|ensures| & \halinline|P =>> loop_invariant (\forall xf;x.min<=xf<x;| \halinline|P[x =>> xf])| \\
\halinline|requires|  & \halinline|P =>> loop_invariant (\forall xf;x<=xf<x.max;| \halinline|P[x =>> xf])| \\
\halinline|context|  & \halinline|P =>> loop_invariant (\forall xf;x.min<=xf<x.max;| \halinline|P[x =>> xf])| \\ \hline \hline
\textbf{With reductions} & For loop nests with reductions, we only use their associated reduction invariant (denoted by \halinline|I|). \\ \hline
Reduction dimensions \halinline|r| & If we passed a previous reduction dimension (\halinline|rprev|), replace that reduction variable with its min: \halinline|I =>> I[rprev =>> rprev.min]| and store that we passed a new reduction invariant. We use the, possibly updated, reduction invariant both here as a loop invariant and store it for the outer loops. \\ \hline
Non-reduction dimension \halinline|x| & First we transform the reduction invariant \halinline|I| into a precondition \halinline|P| and a postcondition \halinline|Q|. \\
& If we have not yet passed a reduction dimension, we look up the next reduction invariant \halinline|rnext| and construct \halinline|P = I| and \halinline|Q = I[rnext =>> rnext + 1]|. \\

& When we have passed reduction dimensions, get the last reduction invariant \halinline|rprev| and construct:
\halinline|P = I[rprev ==> rprev.min]| and \halinline|Q = I[rprev ==> rprev.max]]|. \\
& Similar to the case without reductions, we quantify the invariant over \halinline|x| and store it for the remaining outer loops: \halinline|I =>> (\forall xf;x.min<=xf<x.max; I[x =>> xf])| \\ \hline
\halinline|parallel(x)| & Add the constructed P as a precondition and Q as a postcondition to the parallel block. \\ \hline
\halinline|serial(x)| & Similar to the case without reductions, but for \halinline|P| and \halinline|Q|. \\
& \halinline|loop_invariant (\forall xf;x<=xf<x.max;| \halinline|P[x =>> xf])| \\
& \halinline|loop_invariant (\forall xf;x.min<=xf<x.max;| \halinline|Q[x =>> xf])| \\ \hline \hline
\textbf{Manipulating dimensions} & \\ \hline
\halinline|split(x, xo xi, f)| & Splits \halinline|x|, we replace \halinline|x| and place a guard: \halinline|P =>> xo*f+xi<x.max ==> P[x =>> xo*f+xi] )| \\ \hline
\halinline|fuse(f, x, y)| & Fuse dimension x and y together: \halinline|P =>> P[x =>> f 
\halinline|reorder(x, y)| & Reorders the dimensions \halinline|x| and \halinline|y|. Sets \halinline|x| below \halinline|y| in the loop nest. A dimension can be a reduction.\\ \hline
\textbf{Order of calculation and storage} & \\ \hline
\halinline|f.compute_at(g, x)| & Computes \halinline|f| at the loop of dimension \halinline|x| of the loop nest for function \halinline|g|. The dimensions for \halinline|f| are changed, \haliver makes sure that the annotations respect this. \\ \hline
\halinline|f.store_at(g, x)| & Stores \halinline|f| at the loop of dimension \halinline|x| of the loop nest for function \halinline|g|.  \\ \hline
\end{tabular}
}
\end{table}

\end{document}